\documentclass[12pt]{article}
\usepackage{amsmath}
\usepackage{graphicx}
\usepackage{natbib}
\usepackage{url}
\usepackage{geometry}
\usepackage{placeins}

\graphicspath{ {figures/} }
\usepackage[dvipsnames]{xcolor}
\usepackage[symbol]{footmisc}

\usepackage{amsthm}
\usepackage{amsmath}
\usepackage{amssymb} 
\usepackage{bm}
\usepackage{bbm} 
\usepackage{siunitx}

\usepackage{setspace} 
\usepackage{algorithm}
\usepackage[noend]{algpseudocode}
\newcommand*\Let[2]{\State #1 $\gets$ #2}

\newcommand*\Do[1]{\textbf{do} #1}

\usepackage{tikz}
\newcommand*\circled[1]{\tikz[baseline=(char.base)]{
            \node[shape=circle,draw,inner sep=1pt] (char) {#1};}}
\newcommand{\bmc}[1]{\circled{$\bm{#1}$}}

\def\E{\mathbb{E}}
\def\Var{\mathbb{V}\mathrm{ar}}
\def\Cov{\mathbb{C}\mathrm{ov}}

\def\X{\bm{X}}
\def\Y{\bm{Y}}

\begin{document}
\def\spacingset#1{\renewcommand{\baselinestretch}%
{#1}\small\normalsize}

\spacingset{1.25}

\title{\bf Simultaneous computation of Kendall's tau\\ and its jackknife variance}
\author{Samuel Perreault\thanks{
    The author gratefully acknowledges the support of the Fonds de recherche du Qu\'e{}bec -- Nature et technologies (FRQNT) and the Institut de valorisation des donn\'e{}es (IVADO).} \footnote{e-mail: samuel.perreault@utoronto.ca}\hspace{.2cm}\\
    Department of Statistical Sciences, University of Toronto}

  \maketitle

\begin{abstract}
We present efficient algorithms for simultaneously computing Kendall's tau and the jackknife estimator of its variance.
For the classical pairwise tau, we describe a modification of Knight's algorithm (originally designed to compute only tau) that does so while preserving its $O(n \log_2 n)$ runtime in the number of observations $n$.
We also introduce a novel algorithm computing a multivariate extension of tau and its jackknife variance in $O(n \log_2^p n)$ time.
\end{abstract}


\section{Introduction}

Kendall's tau ($\tau_2$) is a nonparametric measure of pairwise association used in a wide range of scientific fields, notably in survival analysis \citep{Martin/Betensky:2005, Lakhal/Rivest/Abdous:2008}, hydrology \citep{Hamed:2011, Ma/al:2013, Lebrenz/Bardossy:2017} and risk management \citep{McNeil/Frey/Embrechts:2015}.
It is a special case of a much less known multivariate measure ($\tau_p$), generally attributed to \cite{Joe:1990}, which found usage, \emph{e.g.}, for aggregating expert opinions \citep{Jouini/Clemen:1996} and detecting change-points in time series \citep{Quessy/Said/Favre:2013}.
In this paper, we present efficient algorithms for concurrently computing the empirical estimator of $\tau_p$ ($p \geqslant 2$) and the jackknife estimator of its variance.
As they ultimately computes concordance counts (see below), the algorithms can also be used to compute other concordance measures, \emph{e.g.}, Gini's $\gamma$ and Somers' $D$, and their jackknife variance; see \cite{Newson:2006} for more details and other examples.

At its core, $\tau_p$ measures the concordance between pairs of observations, here denoted $c^*(\bm{x},\bm{y}) := \mathbbm{1}(\bm{x}<\bm{y}) + \mathbbm{1}(\bm{x}>\bm{y})$, $\bm{x},\bm{y} \in \mathbb{R}^p$, where $\mathbbm{1}(\bm{x}<\bm{y}) = \prod_{i=1}^p \mathbbm{1}(x_i<y_i)$.
For a random vector $\bm{X} := (X_i)_{i=1}^p$ with continuous distribution $F$, let, following \cite{Joe:1990} and \cite{Jouini/Clemen:1996},
\begin{equation*} \label{eq:tau}
	\tau_p := \frac{2^{p-1} \E\{c^*(\bm{X},\bm{X}')\} - 1}{2^{p-1} - 1} = \frac{2^{p} \int_{\mathbb{R}^p} F(\bm{x}) \mathrm{d} F(\bm{x}) - 1}{2^{p-1} - 1} \;,
\end{equation*}
which covers Kendall's original measure as special case ($p=2$).
One usually estimates $\tau_p$ based on the concordance count $c := (1/2)\sum\nolimits_{i=1}^n c_i$, where $c_i := \sum_{j=1,j\neq i}^n c^*(\bm{X}_i,\bm{X}_j)$ counts the observations with which $\bm{X}_i$ is concordant, or the discordance count $d := (1/2)\sum\nolimits_{i=1}^n d_i = \binom{n}{2} - c$, where $d_i := n - 1 - c_i$; further let $\bm{c} := (c_i)_{i=1}^n$ and $\bm{d} := (d_i)_{i=1}^n$.
The corresponding estimator of $\tau_p$ is the U-statistic \citep{Hoeffding:1948} of order two
\begin{align} \label{eq:tau-hat}
	\hat\tau_p := \frac{2^{p-1} c/\binom{n}{2} - 1}{2^{p-1}-1} = \frac{1 - 2^{p-1} d/\binom{n}{2}}{2^{p-1}-1}\;,
\end{align}
with underlying kernel $h(\bm{x},\bm{y}) := \{2^{p-1} c^*(\bm{x},\bm{y}) - 1\}/(2^{p-1} - 1)$.

Traditionally, inference for $\tau_p$ is based on the asymptotic normality of U-statistics, whose derivation usually involves the projection of $(\hat\tau_p - \tau_p)$ onto the space of linear functions, denote it $H_n := (2/n) \sum_{i=1}^n g_i$, where $g_i := g(\bm{X}_i)$ with $g(\bm{x}) := \E\{h(\bm{x},\bm{X})\} - \tau_p$.
Notably, \cite{Sen:1972} showed, in a more general framework and under mild conditions, that in the presence of serial dependence, $\sqrt{n}(\hat{\tau}_p - \tau_p) \rightsquigarrow \mathcal{N}(0,\sigma_p^2)$ as $n \to \infty$, where $\sigma_p^2 := \Var(\sqrt{n}H_n) = 4 \zeta_0 + 8 \sum\nolimits_{j=1}^\infty \zeta_j$, with $\zeta_j := \Cov(g_1,g_{1+j})$ such that $\zeta_j = 0$ ($j \geqslant 1$) in the absence of serial dependence.
This was used by \cite{Ferguson/Genest/Hallin:2000} to construct a measure of serial dependence based on $\tau_2$.

There exist several methods for estimating $\sigma_p^2$ from independent observations: a plug-in estimator proposed by \citet[Appendix]{Genest/Neslehova/BenGhorbal:2011} for the case $p=2$; jackknife estimation \citep{Arvesen:1969,Callaert/Veraverbeke:1981,Rublik:2016}; or implicit estimation via bootstrap methods (\citeauthor{Arcones/Gine:1992}, \citeyear{Arcones/Gine:1992}; \citeauthor{Chen:2018}, \citeyear{Chen:2018}).
We focus here on an extension of the (rescaled) jackknife estimator appearing in \cite{Chen:2018},
\begin{equation} \label{eq:g-hat}
	 \hat{\sigma}_p^2 := \frac{4}{n} \left( \sum_{i=1}^n \hat{g}_i^2 + 2 \sum_{j=1}^{m} \sum_{i=1}^{n-j} \hat{g}_i\hat{g}_{i+j} \right),\ \hat{g}_i := \frac{2^{p-1} c_i/(n-1) - 1}{2^{p-1}-1} - \hat\tau_p\;,
\end{equation}
which further accounts for serial dependence up to lag $m$. In practice, $m$ is generally not known and weighting/truncation schemes similar to those in \cite{Andrews:1991} may be used to favour a stable asymptotic behaviour.

To see how $\hat\sigma_p^2$ is related to the jackknife estimator, denote it $\tilde\sigma_p^2$, note that $\tilde\sigma_p^2 = (n-1)\sum_{i=1}^n (\hat\tau_p^{(i)} - \hat\tau_p)^2$, where $\hat\tau_p^{(i)}$ is the replicate of $\hat{\tau}_p$ obtained by discarding the $i$th observation, and that $\hat{g_i} = (n-2)(\hat\tau_p - \hat\tau_p^{(i)})/2$.
It follows from these identities that, with $m=0$ in \eqref{eq:g-hat}, $\hat\sigma_2^2 = (n-2)^2\tilde\sigma_p^2/\{n(n-1)\}$. See \citet[Section~3.3]{Chen:2018} and references therein for more details.
The extension ($m \geqslant 0$) is obtained by summing over $\sum_{i=1}^n \sum_{j=i}^{i+m} \sum_{k \neq i} \sum_{\ell \neq j}$ as opposed to $\sum_{i=1}^n \sum_{k \neq i} \sum_{\ell \neq i}$ in (18) of the latter author.

Since $\hat{\sigma}_p^2$ can be computed in linear time given $\bm{\hat{g}} := (\hat{g}_i)_{i=1}^n$, we focus on the computation of $\bm{c}$ or $\bm{d}$.
A similar strategy was employed by \cite{Knight:1966} to develop the now standard $O(n \log_2 n)$ computing algorithm for $\hat\tau_2$, building on a quadratic method described by \cite{Kendall:1948}.
Since then, variations of Knight's algorithm were proposed by \cite{Abrevaya:1999}, \cite{Christensen:2005} and \cite{Newson:2006}. 
Newson's \texttt{Stata} implementation seems to have been the first, and up to very recently, the only one allowing the computation of both $\hat\tau_2$ and $\hat{\sigma}_2^2$ in $O(n \log_2 n)$ time.

The idea underlying Kendall's and Knight's algorithms are reviewed in Section~\ref{sec:pairwise-tau}, where we also describe an extension of the latter for computing $\hat\sigma_2^2$.
This algorithm extends Knight's algorithm more directly than that of \cite{Newson:2006}, which is based on balanced binary search trees; it was independently sketched by \cite{Dufey:2020} \citep[see also][]{Raymaekers/Dufey:2022} and the present author (unpublished work from \cite{Perreault:2020}).
In Section~\ref{sec:multivariate-tau}, we introduce a novel $O(n \log_2^p n)$ algorithm for computing $\hat\tau_p$ and $\hat\sigma_p^2$ when $p \geqslant 2$, thus outperforming the $\Theta(n^2)$ naive algorithm asymptotically.
In Section~\ref{sec:benchmark}, we present a numerical study investigating the performance of the algorithms in finite samples.

Appendices available online provide pseudocode for the algorithms (Appendix~A), the analysis of the new algorithm's runtime (Appendix~B), detailed results of the numerical study (Appendix~C), and an illustrative application in which we define a multi-lag autocorrelation measure based on $\hat\tau_p$ to quantify the temporal persistence of environmental phenomena (Appendix~D). Throughout, we follow \cite{Cormen/al:2014} and use $\Theta(\cdot)$ and $O(\cdot)$ to describe asymptotically tight bounds and asymptotic upper bounds, respectively, for the time complexity of algorithms.

\section{Computation of $\hat\tau_2$ and $\hat\sigma_2^2$} \label{sec:pairwise-tau}

\subsection{Kendall's method} \label{sec:kendall}

Consider first the case $p=2$, and let $\X := (X_i)_{i=1}^n$ and $\Y := (Y_i)_{i=1}^n$ be the two vectors of observations. 
The computation of $\hat{\tau}_2$ is closely linked to the $O(n^2)$ \emph{exchange sort} algorithm, which sorts a vector by scanning it and swapping any two adjacent numbers not in order until no further swaps are possible \citep{Knight:1966}.
\cite{Kendall:1948} noted that, given $X_1 < \dots < X_n$, \emph{exchange sort} performs exactly $d$ swaps when sorting $\Y$, as each swap removes exactly one discordance.
Thus, one can compute $\hat{\tau}_2$ by first sorting $\bm{X}$ and re-ordering $\bm{Y}$ accordingly, and then sorting $\Y$ with \emph{exchange sort}, keeping track of the swap count.
This is depicted in Figure~A.1.

\subsection{Knight's method} \label{sec:knight}

\cite{Knight:1966} improved on Kendall's method by noting that \emph{merge sort} algorithms, whose runtime is $O(n \log_2 n)$, allow tracking the number of swaps that \emph{exchange sort} would have required.
Assuming for simplicity that $n=2^K$ for some $K \in \mathbb{N}$, Knight's algorithm first partitions $\Y$ into $2^{K-0}$ subvectors of length $2^0 = 1$ and then recursively merges them in pairs, keeping the newly created subvectors sorted.
After one recursive step, the algorithm produces $2^{K-1}$ sorted subvectors of size $2^1$; after two, it produces $2^{K-2}$ sorted subvectors of size $2^2$; and so on until the full vector is sorted.

An $O(n \log_2 n)$ runtime is achieved by exploiting the fact that any two vectors to be merged, say $\bm{a} = (a_i)_{i=1}^m$ and $\bm{b} = (b_i)_{i=1}^m$ ($m \in \mathbb{N}$), are already sorted.
As depicted in Figure~A.2, a typical \emph{merge} phase constructs the sorted output, say $\bm{v}$, recursively by initialising $i \leftarrow 1,\ j \leftarrow 1$ and then alternating between the updates $v_{i + j - 1} \leftarrow \min(a_i, b_j)$ and $(i,j) \leftarrow (i,j) + (z_{ij}, 1-z_{ij})$, where $z_{ij} = \mathbbm{1}(a_i < b_j)$, until either $a_m$ or $b_m$ is selected. The remaining entries (all from the same vector, $\bm{a}$ or $\bm{b}$) are then selected in their original order.

To deduce the number of swaps that \emph{exchange sort} would require to sort $(\bm{a},\bm{b})$, note that after $(i-1)$ and $(j-1)$ components have been selected from $\bm{a}$ and $\bm{b}$, respectively, there remains only to sort $(a_i,\dots,a_m,b_j,\dots,b_m)$.
At this stage, $a_i$ is selected precisely when $b_{j-1} < a_i < b_j$, and \emph{exchange sort} would thus need to swap it $j-1$ times (with $b_1,\dots,b_{j-1}$).
Similarly, $b_j$ is selected if and only if $a_{i-1} < b_j < a_i$, which would trigger \emph{exchange sort} to swap it $m-i+1$ times (with $a_i,\dots, a_m$).

\subsection{Jackknife variance via Knight's method} \label{sec:knight-extended}

The key step in devising an algorithm for computing $\hat\sigma_2^2$ in $O(n \log_2 n)$ time is to note, reasoning similarly as before, that $d_i$ corresponds to the number of swaps involving $Y_i$ during \emph{exchange sort}, 
and that Knight's algorithm can be extended to compute the individual swap counts $\bm{d}$.
To see how, consider again merging two sorted vectors $\bm{a} = (a_i)_{i=1}^m$ and $\bm{b} = (b_i)_{i=1}^m$ ($m \in \mathbb{N}$) using its \emph{merge} phase, as in Section~\ref{sec:knight}.
Note that the swap count given therein (\emph{i.e.}, $j-1$ when $a_i$ is selected and $m-i+1$ when $b_j$ is selected) is in fact the individual swap count associated with the selected entry.
To properly cumulate individual swap counts across multiple \emph{merge} phases, it suffices to keep track of the original position of each observation as $\bm{Y}$ is being sorted, which can be achieved without altering the algorithm's time complexity.
Algorithms~A.1--A.3 provide pseudocode for this extension of Knight's algorithm; they expand upon code from the \texttt{R} package \texttt{pcaPP} \citep{Filzmoser/Fritz/Kalcher:2018} implementing Knight's original algorithm.

\section{Computation of $\hat\tau_p$ and $\hat\sigma_p^2$} \label{sec:multivariate-tau}

Knight's algorithm is not directly applicable when $p > 2$, but it can still be used to compute $\hat\tau_3$, since $\hat\tau_3 = (\hat\tau_{1,2} + \hat\tau_{2,3} +\hat\tau_{1,3})/3$, where $\hat\tau_{i,j}$ is the pairwise measure associated with variables $i$ and $j$ \citep{Nelsen:1996}. 
Thus, provided that $X_{11} < \cdots < X_{n1}$ in the initial state, $\hat\tau_3$ can be computed by recursively sorting the rows of $(\bm{X}_{i})_{i=1}^n$ with respect to the values in its second, third and first columns.
Because the initial and final states are the same, two rows are swapped together either zero times (when concordant), or twice (when discordant). Consequently, and in a similar notation, $\bm{d} = (\bm{d}_{1,2} + \bm{d}_{2,3} + \bm{d}_{1,3})/2$, which allows computing $\hat\sigma_3^2$ in $O(n \log_2 n)$ time.

The recursive relation linking $\hat\tau_p$ to its lower-dimensional analogues when $p=3$ is a special case of that given by \citet[Proposition~1]{Genest/Neslehova/BenGhorbal:2011}.
Valid only for $p$ odd, this latter requires computing $\hat\tau_{I}$, the $|I|$-variate measure associated with a subset of variables $I$, for all even-sized $I \subset \{1,\dots,p\}$.
Here, we present instead an algorithm that computes $\bm{\hat\tau}_p := (\hat\tau_{\{1,\dots,k\}})_{k=2}^p = (\hat\tau_k)_{k=2}^p$ and its jackknife covariance matrix, denote it $\bm{\hat\Sigma}_p$.
This latter is defined analogously as $\hat\sigma_p^2$ in \eqref{eq:g-hat}, but with $\hat{g}_i$ replaced by the $(p-1)$-dimensional vector associated with $\bm{\hat\tau}_p$.
In Appendix~D, we highlight how $\hat\tau_p$ can be used in the context of time series analysis to measure a new kind of multi-lag autocorrelation; in this case $\bm{\hat\tau}_p$ consists of the autocorrelations with maximum lag ranging from 1 to $p-1$.

Computing $\bm{\hat\tau}_p$ and $\bm{\hat\Sigma}_p$ requires computing, for each dimension $k \in \{2,\dots,p\}$, the vector of discordance counts $\bm{d}^{(k)}$ ($\bm{d}$ defined above \eqref{eq:tau-hat}, with $p=k$) associated with $\hat\tau_k$ and $\hat\sigma_k^2$.
We propose to compute $(\bm{d}^{(k)})_{k=2}^p$ using a divide-and-conquer stategy.
The full procedure is given, in pseudocode, in Algorithm~A.4.
For simplicity, we focus here on the computation of the corresponding global counts, denoted $(d^{(k)})_{k=2}^p$, assuming again that $X_{11} < \dots < X_{n1}$.
For any $\mathcal{I},\mathcal{J} \subseteq \{1,\dots,n\}$ and $k \in \{2,\dots,p\}$, 
let $D_\mathcal{I}^{(k)}$ and $D_{\mathcal{I}\mathcal{J}}^{(k)}$ be the numbers of pairs within $\{\bm{X}_i\}_{i \in \mathcal{I}}$ and between $\{\bm{X}_i\}_{i \in \mathcal{I}}$ and $\{\bm{X}_j\}_{j \in \mathcal{J}}$, respectively, that are concordant only up to dimension $k$, and set $\bm{D}_\mathcal{I} := (D_\mathcal{I}^{(k)})_{k=2}^p$ and $\bm{D}_{\mathcal{I}\mathcal{J}} := (D_{\mathcal{I}\mathcal{J}}^{(k)})_{k=2}^p$.
We now describe how to compute $\bm{D}_{\mathcal{I}}$, from which we get $d^{(k)} = \sum_{r=2}^k D_{\{1,\dots,n\}}^{(r)}$.

Let $\mathcal{I} = \{i,i+1, \dots, i+m-1\} \subseteq \{1,\dots,n\}$. Towards computing $\bm{D}_{\mathcal{I}}$, Algorithm~A.4 first checks whether $m:=|\mathcal{I}|$ is smaller than some user-specified threshold $n_*$.
If so, $\bm{D}_{\mathcal{I}}$ is computed naively in $\Theta(n_*^2) = \Theta(1)$ operations, and $\mathcal{I}$ is reordered ($\mathcal{I} = \{i_r\}_{r=1}^m$) so that $X_{i_r 2} < X_{i_{r+1}2}$ for all $r < m$.
If not, but $p=2$, the \emph{merge} phase of Knight's algorithm is used to compute $D_{\mathcal{I}}^{(2)}$ and the procedure terminates.
Otherwise, $\mathcal{I}$ is halved into two sets $\mathcal{I}_0$ and $\mathcal{I}_1$ (see Figure~A.3) so that $\bm{D}_{\mathcal{I}} = \bm{D}_{\mathcal{I}_0} + \bm{D}_{\mathcal{I}_1} + \bm{D}_{\mathcal{I}_0\mathcal{I}_1}$. 
Thus, $n_*$ is chosen so that it is cheaper to naively compute $\bm{D}_{\mathcal{I}}$ than it is to proceed by further halving $\mathcal{I}$; we use $n_*=10$.
Recursive calls are used to compute $\bm{D}_{\mathcal{I}_{0}}$ and $\bm{D}_{\mathcal{I}_{1}}$, after which it remains only to deal with $\bm{D}_{\mathcal{I}_0\mathcal{I}_1}$.

When $|\mathcal{I}_0| \cdot |\mathcal{I}_1| > n_*^2$, 
$\bm{D}_{\mathcal{I}_0\mathcal{I}_1}$ is computed as follows; otherwise it is computed naively in $\Theta(1)$.
At this stage, by design, $X_{i1} < X_{j1}$ for all $(i,j) \in \mathcal{I}_0 \times \mathcal{I}_1$ and the index sets were reordered so that $X_{i_r 2} < X_{i_{r+1} 2}$ for all $i_r,i_{r+1} \in \mathcal{I}_\delta$ and $\delta \in \{0,1\}$ (during the computation of $\bm{D}_{\mathcal{I}_{0}}$ and $\bm{D}_{\mathcal{I}_{1}}$).
The sets $\mathcal{I}_0$ and $\mathcal{I}_1$ are further split to form a new partition $\mathcal{I}_{\delta}^{\gamma} = \{ i \in \mathcal{I}_\delta : \gamma_i = \gamma\}$, $\delta,\gamma \in \{0,1\}$, where $\gamma_i = \mathbbm{1}(X_{i2} < s)$ with $s$ chosen such that the count of pairs from diagonally positioned sets, denote it $\beta := |\mathcal{I}_{0}^0|\cdot |\mathcal{I}_{1}^1| + |\mathcal{I}_{1}^0|\cdot |\mathcal{I}_{0}^1|$, is maximal; see Figure~A.3 for an example. 
This maximises the number of pairs of observations for which a conclusion is reached: the pairs in $\mathcal{I}_{0}^0 \times \mathcal{I}_{1}^1$ are concordant on their first two dimensions, meaning that $D_{\mathcal{I}_{0}^0\mathcal{I}_{1}^1}^{(2)} = 0$, and those in $\mathcal{I}_{1}^0 \times \mathcal{I}_{0}^1$ are necessarily discordant and need not be further investigated, that is, $\bm{D}_{\mathcal{I}_{0}^1\mathcal{I}_{1}^0} = (|\mathcal{I}_{0}^1| \cdot |\mathcal{I}_{1}^0|, \bm{0})$.

The optimal split is found in linear time using a slight modification of the \emph{merge} procedure described in Section~\ref{sec:knight} and depicted in Figure~A.2.
The new procedure still records the reordering of $\mathcal{I}$, say $\{j_r\}_{r=1}^m$, for which $X_{j_r 2} < X_{j_{r+1} 2}$ for all $r<m$ (to be applied at termination of Algorithm~A.4), but also the value of $\beta$ for each possible split value among $(X_{i 2})_{i \in \mathcal{I}}$.
Because $\mathcal{I}_0$ and $\mathcal{I}_1$ are split using the same value $s$ (on the y-axis in Figure~A.3), we have that $\bm{D}_{\mathcal{I}_0\mathcal{I}_1} = \bm{D}_{\mathcal{I}_0^0\mathcal{I}_1^0} + \bm{D}_{\mathcal{I}_0^1\mathcal{I}_1^0} + \bm{D}_{\mathcal{I}_0^0\mathcal{I}_1^1} + \bm{D}_{\mathcal{I}_0^1\mathcal{I}_1^1}$, and it remains only to compute $\bm{D}_{\mathcal{I}_{0}^0\mathcal{I}_{1}^0}$, $\bm{D}_{\mathcal{I}_{0}^1\mathcal{I}_{1}^1}$ and $(D_{\mathcal{I}_{0}^0\mathcal{I}_{1}^1}^{(k)})_{k=3}^{p}$, which is done by Algorithm~A.5.

To avoid redundancy and cumbersome notation, we omit the description of Algorithm~A.5 here, as it operates similarly to the latter part of Algorithm~A.4.
It is worth pointing out, however, that when it is invoked to compute quantities of the form $(D_{\mathcal{I}_{0}^0\mathcal{I}_{1}^1}^{(k)})_{k=\ell}^p$ for some $\ell \in \{3,\dots,p\}$, the indices in $\mathcal{I}_{\delta}^\delta$ ($\delta = 0,1$) might not be ordered such that $X_{i_{r} \ell} < X_{i_{r+1} \ell}$ for all $i_r, i_{r+1} \in \mathcal{I}_{\delta}^\delta$. 
As this condition is crucial for the algorithm to function properly, these calls necessitate an extra sorting step. 
Also note that, for $\ell = p$, Algorithm~A.5 is a terminal operation consisting, in addition to the sorting step, of a single \emph{merge} operation.
These considerations play a role in the runtime analysis provided in Appendix~B, which shows that the algorithm runs in $\Theta(n \log_2 n)$ time when the components of $\bm{X}$ are co-monotone, and in $O(n \log_2^p n)$ more generally.

\section{Benchmarking} \label{sec:benchmark}

We now investigate the runtime performance in finite-samples of the extended Knight's algorithm (KE, Algorithm~A.1) outlined in Section~\ref{sec:knight-extended} and the divide-and-conquer algorithm (DAC, Algorithm~A.4) outlined in Section~\ref{sec:multivariate-tau}.
As benchmarks for evaluation, we also consider their naive $\Theta(n^2)$ alternative (BF, for brute force), which explicitly considers all pairs of observations, as well as Knight's original algorithm (KO), which only computes $\hat\tau_2$.
All algorithms were implemented in \texttt{C++}; for KO, we used the implementation of the \texttt{pcaPP} package in \texttt{R} \citep[\texttt{cor.fk}]{Filzmoser/Fritz/Kalcher:2018}.
The code is available as online supplementary material.

\begin{figure}[t]
\centering
\includegraphics[width=.975\textwidth]{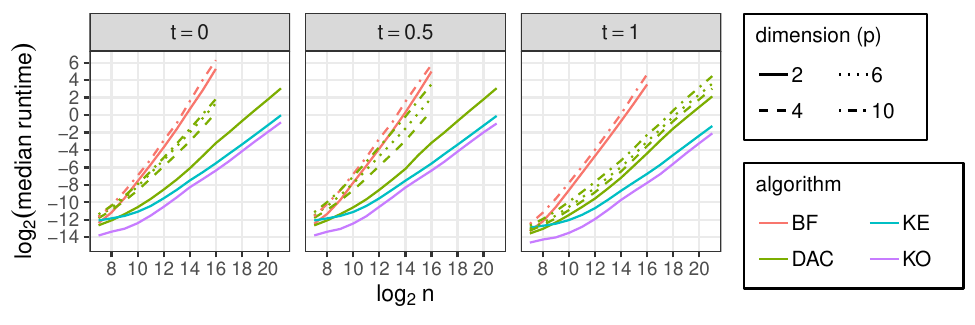}
\caption{Median runtime (in seconds) over $100$ datasets of Knight's original (KO) algorithm, the extension (KE) from Section~\ref{sec:knight-extended}, the divide-and-conquer algorithm (DAC) from Section~\ref{sec:multivariate-tau} and their naive alternative (BF) for computing $\bm{\hat\tau}_p$ and (except for KO) $\bm{\hat\Sigma}_p$, as defined in Section~\ref{sec:multivariate-tau}, for each $p \in \{2,4,6,10\}$ (when applicable). The data are equicorrelated normal random vectors with correlation $\rho_t = \sin(\pi t/2)$, $t \in \{0,0.5,1\}$.}
\label{fig:bench}
\end{figure}

To analyse the runtimes, we generated $10$-dimensional normal samples with Pearson correlation matrix given by $\bm{R} = (1-\rho) \bm{I} + \rho$. In this case, $\rho$ and $\tau_2$ are related by the identity $\rho = \sin(\pi \tau_2 / 2)$ \citep{Esscher:1924}. For various sample sizes $n \in \{2^i\}_{i=7}^{21}$ and each value of $t \in \{i/4\}_{i=0}^4$, we generated $100$ datasets with $\rho_t := \sin(\pi t / 2)$. For each sample, we computed $\bm{\hat\tau}_p$ and $\bm{\hat\Sigma}_p$ (as in Section~\ref{sec:multivariate-tau}) with DAC for each $p \in \{2,4,6,10\}$, and with BF for $p \in \{2,10\}$. KO and KE are by design restricted to the case $p=2$.
For smaller runtimes, we repeated each computation several times (at most 100) and used the median as official runtime.
Longer computations were performed only once.
The minimum, median and maximum runtimes for each algorithm are reported on the log-log scale in Figure~C.1.
The median runtimes for $\rho_t$, $t \in \{0, 0.5, 1\}$, with which the minimum and maximum roughly agree, are reproduced in Figure~\ref{fig:bench}.

As Figure~\ref{fig:bench} illustrates, the curves corresponding to BF are all nearly linear with slopes of $\sim$2, which is consistent with a $\Theta(n^2)$ runtime. 
The results also confirm that KO and KE exhibit similar behavior, with runtimes that are nearly linear for large sample sizes, as indicated by their slopes of $\sim$1 on the log-log scale. Furthermore, they suggest that the additional steps required to compute $\hat{\sigma}_2^2$ inflate the runtime by a factor smaller than four for small sample sizes and smaller than two for larger ones. 

For $t<1$, DAC's curves have slopes that initially increase with $n = 2^k$ to attain a maximum value around $k=16$ and then decrease to a value close to $1 + p\{\log_2(k) - \log_2(k-1)\}$, as one would expect from an algorithm with a $\Theta(n \log_2^p n)$ runtime; see Figure~C.2 for extra simulations with larger values of $k$.
In general, increasing $t$ reduces the probability of discordance, which provides fewer opportunities for DAC to reach definitive conclusions and, for $t$ away from $1$ and $p > 2$, triggers many more extra sorting steps, thus increasing DAC's runtime.
In the worst case, for $t=0.5$ and $p=10$, the runtimes of DAC and BF are comparable for smaller values of $n$, but eventually depart significantly; at $n=2^{16}$, DAC has a median runtime of $\sim$11.6, while that of BF is $\sim$54.25.
As $t$ approaches one, the probability that DAC encounters large sets that are already well-ordered becomes more significant, making its runtime drop sharply.
With $t=1$, DAC's curves are nearly parallel and their slopes slowly approach that of KO and KE, corroborating a $\Theta(n \log_2 n)$ runtime for all $p$.

Finally, note that for $t < 1$ and $n$ fixed, the median runtime of DAC should reach an upper bound as $p \to \infty$, since in this case there must exist $k$ such that $\hat\tau_p = 0$ for all $p \geqslant k$, saving DAC from exploring beyond the $k$th dimension. In Figure~\ref{fig:bench}, this is most apparent when $t=0$, as the probability that $\hat\tau_p = 0$ is larger in this case; see also Figure~D.5.
Relatedly, re-labelling variables to prioritise discordance on initial dimensions (\emph{e.g.}, based on pairwise taus) can reduce DAC's runtime, although this should be negligible when the variables are exchangeable, as they are here.

\subsection*{Acknowledgements}

The author wishes to thank Professors N.\ Reid and P.\ Brown, and anonymous referees for their valuable comments on earlier versions of the paper. This work was supported by the \emph{Fonds de recherche du Qu{\'e}bec -- Nature et technologies} and the \emph{Institut de valorisation des donn{\'e}es}.



\singlespacing
\bibliographystyle{elsarticle-harv}

\newpage
\spacingset{1.25}
\appendix
\renewcommand{\thealgorithm}{\Alph{section}.\arabic{algorithm}}
\setcounter{figure}{0}
\renewcommand{\thefigure}{\thesection.\arabic{figure}}  
\counterwithin{algorithm}{section}
\counterwithin{figure}{section}

\newpage
\section{Algorithms} \label{app:algorithms}

\begin{figure}[H]
\singlespace
\centering
\includegraphics[scale=1]{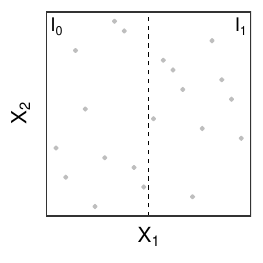}
\hspace{1cm}
\includegraphics[scale=1]{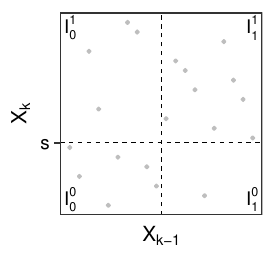}
\caption{Depiction of the sets $\mathcal{I}_\delta$ ($\delta \in \{0,1\}$, left panel) and $\mathcal{I}_\delta^\gamma$ ($\delta,\gamma \in \{0,1\}$, right panel) formed during an iteration of the divide-and-conquer algorithm outlined in Section~3. Each point in the panels represents an observation, and the parameter $s$ is selected to optimise the count of pairs from diagonally positioned sets.} \label{fig:dac-toy}
\end{figure}

\begin{itemize}

	\item \textbf{Figures~\ref{fig:exchange-sort}--\ref{fig:merge-sort}:} Depiction of Kendall's and Knights's methods discussed in Sections~2.1 and 2.2.

	\item \textbf{Figure~\ref{fig:dac-toy}:} Depiction of the sets $\mathcal{I}_\delta$ ($\delta \in \{0,1\}$) and $\mathcal{I}_\delta^\gamma$ ($\delta,\gamma \in \{0,1\}$) formed during an iteration of the divide-and-conquer algorithm outlined in Section~3.

	\item \textbf{Algorithms~\ref{alg:merge-sort}--\ref{alg:merge-sort-single}:} Pseudocode for the extended Knight's algorithm, presented in Section~2.3. It is based on the \texttt{C++} code underlying the function \texttt{cor.fk} of the \texttt{R} package \texttt{pcaPP} \citep{Filzmoser/Fritz/Kalcher:2018}.
Given $\X$ and $\Y$ as in Section~2, the general idea is to recursively split $\Y$ (or subvectors thereof) into two subvectors until these are of length less than $10$ (Algorithm~\ref{alg:merge-sort}); these smaller subvectors are then sorted using so-called \emph{insertion sort} (Algorithm~\ref{alg:insertion-sort}) and recursively merged (Algorithm~\ref{alg:merge-sort-single}).

	\item \textbf{Algorithms~\ref{alg:dac-within}--\ref{alg:dac-between}:} Pseudocode for the divide-and-conquer algorithm presented in Section~3. These latter involve subroutines (\textsc{Conquer}, \textsc{Conquer*}, \textsc{Sort}, \textsc{Merge*}, \textsc{Merge**}, \textsc{FindSplitSort} and \textsc{FindSplit}), given as part of the \texttt{C++} code provided as online supplementary material.
\end{itemize}

\pagebreak

\begin{figure}[H]
\singlespace
\centering
\fontsize{11.45pt}{12pt}\selectfont
$\begin{array}{c|c}
\X' & \Y' \\
\hline
\circled{2} & 2\\
\circled{3} & 1\\
\circled{1} & \bm{3}\\
\circled{4} & 5\\
\circled{5} & 4\\
\hline
\multicolumn{2}{c}{\vspace{-.4cm}}\\
\multicolumn{2}{c}{\text{Step 0}}
\end{array}$
\quad
$\begin{array}{c|c||c}
\X & \Y & \bm{d}^0 \\
\hline
1 & \circled{$\bm{3}$} & 0 \\
2 & \circled{2} & 0 \\
3 & 1 & 0 \\
4 & 5 & 0 \\
5 & 4 & 0 \\
\hline
\multicolumn{3}{c}{\vspace{-.4cm}}\\
\multicolumn{3}{c}{\text{Step 1}}
\end{array}$
\quad
$\begin{array}{c||c}
\Y^1 & \bm{d}^1 \\
\hline
2 & 1 \\
\circled{$\bm{3}$} & 1 \\
\circled{1} & 0 \\
5 & 0 \\
4 & 0 \\
\hline
\multicolumn{2}{c}{\vspace{-.4cm}}\\
\multicolumn{2}{c}{\text{Step 2}}
\end{array}$
\quad
$\begin{array}{c||c}
\Y^2 & \bm{d}^2 \\
\hline
2 & 1 \\
1 & 1 \\
\circled{$\bm{3}$} & 2 \\
\circled{5} & 0 \\
4 & 0 \\
\hline
\multicolumn{2}{c}{\vspace{-.4cm}}\\
\multicolumn{2}{c}{\text{Step 3}}
\end{array}$
\quad
$\begin{array}{c||c}
\Y^2 & \bm{d}^2\\
\hline
2 & 1 \\
1 & 1 \\
\bm{3} & 2 \\
\circled{5} & 0 \\
\circled{4} & 0 \\
\hline
\multicolumn{2}{c}{\vspace{-.4cm}}\\
\multicolumn{2}{c}{\text{Step 4}}
\end{array}$
\quad
$\begin{array}{c||c}
\Y^3 & \bm{d}^3 \\
\hline
\circled{2} & 1 \\
\circled{1} & 1 \\
\bm{3} & 2 \\
4 & 1 \\
5 & 1 \\
\hline
\multicolumn{2}{c}{\vspace{-.4cm}}\\
\multicolumn{2}{c}{\text{Step 5}}
\end{array}$
\ \dots\
$\begin{array}{c||c}
\Y^4 & \bm{d}^4 \\
\hline
1 & 2 \\
2 & 2 \\
\bm{3} & 2 \\
4 & 1 \\
5 & 1 \\
\hline
\multicolumn{2}{c}{\vspace{-.4cm}}\\
\multicolumn{2}{c}{\text{Output}}
\end{array}$
\caption{\small Depiction of Kendall's method: sort $\X'$  (Step 0), and then apply \emph{exchange sort} to $\Y$ (Step 1--5). Three steps resulting in no swap were omitted (Steps 6, 7 and 8). 
$\bm{Y}^k$ is the version of $\Y = \Y^0$ that results from $k$ swaps; circled entries are those being compared and reordered, if necessary; those in bold track the position of $Y_1$; and $\bm{d}^k$ records the entry-specific swap counts after $k$ swaps.
In particular, $\bm{d}^4$ is a permutation of $(d_1,\dots,d_5)$, and $d = (d_1+ \dots +d_5)/2 = 4$ swaps are required to sort $\bm{Y}$ (Steps~1, 2, 4 and 5).
} \label{fig:exchange-sort}
\end{figure}

\bigskip

\begin{figure}[H]
\singlespace
\fontsize{11pt}{12pt}\selectfont
\centering
$\begin{array}{c|c}
\bm{a} & \bm{b}\\
\hline
\bmc{1} & \circled{2}\\
4 & 5\\
6 & 7\\
8 & 9\\
\hline
\multicolumn{2}{c}{\vspace{-.4cm}}\\
\multicolumn{2}{c}{v_1 \leftarrow 1}
\end{array}$
\quad
$\begin{array}{c|c}
\bm{a} & \bm{b}\\
\hline
1 & \bmc{2}\\
\circled{5} & 4\\
6 & 8\\
7 & 9\\
\hline
\multicolumn{2}{c}{\vspace{-.4cm}}\\
\multicolumn{2}{c}{v_2 \leftarrow 2}
\end{array}$
\quad
$\begin{array}{c|c}
\bm{a} & \bm{b}\\
\hline
1 & 2\\
\circled{5} & \bmc{4}\\
6 & 8\\
7 & 9\\
\hline
\multicolumn{2}{c}{\vspace{-.4cm}}\\
\multicolumn{2}{c}{v_3 \leftarrow 4}
\end{array}$
\quad
$\begin{array}{c|c}
\bm{a} & \bm{b}\\
\hline
1 & 2\\
\bmc{5} & 4\\
6 & \circled{8}\\
7 & 9\\
\hline
\multicolumn{2}{c}{\vspace{-.4cm}}\\
\multicolumn{2}{c}{v_4 \leftarrow 5}
\end{array}$
\quad
$\begin{array}{c|c}
\bm{a} & \bm{b}\\
\hline
1 & 2\\
5 & 4\\
\bmc{6} & \circled{8}\\
7 & 9\\
\hline
\multicolumn{2}{c}{\vspace{-.4cm}}\\
\multicolumn{2}{c}{v_5 \leftarrow 6}
\end{array}$
\quad
$\begin{array}{c|c}
\bm{a} & \bm{b}\\
\hline
1 & 2\\
5 & 4\\
6 & \circled{8}\\
\bmc{7} & 9\\
\hline
\multicolumn{2}{c}{\vspace{-.4cm}}\\
\multicolumn{2}{c}{v_6 \leftarrow 7}
\end{array}$
\quad
$\begin{array}{c|c}
\bm{a} & \bm{b}\\
\hline
1 & 2\\
5 & 4\\
6 & \bmc{8}\\
7 & 9\\
\hline
\multicolumn{2}{c}{\vspace{-.4cm}}\\
\multicolumn{2}{c}{v_7 \leftarrow 8}
\end{array}$
\quad
$\begin{array}{c|c}
\bm{a} & \bm{b}\\
\hline
1 & 2\\
5 & 4\\
6 & 8\\
7 & \bmc{9}\\
\hline
\multicolumn{2}{c}{\vspace{-.4cm}}\\
\multicolumn{2}{c}{v_8 \leftarrow 9}
\end{array}$
\caption{\small Depiction of the \emph{merge} phase of \emph{merge sort} from Knight's algorithm.
Two sorted vectors $\bm{a},\bm{b} \in \mathbb{R}^4$ are merged into a new vector $\bm{v} \in \mathbb{R}^8$.
The entries circled are those being compared at a given step, and those in bold are the selected ones.}
\label{fig:merge-sort}
\end{figure}

\bigskip

\pagebreak
\singlespace
\begin{algorithm}[H]
  \caption{\emph{Merge sort} algorithm keeping the swap counts.} \label{alg:merge-sort}
  \begin{algorithmic}[1]
    \Require{$\bm{y}, \bm{x}$: numeric vectors of length $m, m$.}  \Comment{observations and storage}
    \Require{$\bm{q}, \bm{p}, \bm{d}$: integer vectors of lengths $m, m, n$.} \Comment{ids, storage and swap counts}
    \item[] This function is recursive. It is assumed that, in total, $n$ observations $(y_1,\dots,y_n)$ are available, that $m = n$ when the procedure is first called and $m \leqslant n$ in general; $\bm{q}$ is initialized with $(1,\dots,n)$; $\bm{d}$ is initialized with $n$ zeros. \textsc{InsertionSort} and \textsc{Merge} are given by Algorithms~\ref{alg:insertion-sort}-\ref{alg:merge-sort-single}.

    \Statex
    \Procedure{MergeSort}{$\bm{y}, \bm{x}, \bm{q}, \bm{p}, \bm{d}$}
    \If{$m < 10$}
    \Do{\textsc{InsertionSort}($\bm{y}, \bm{q}, \bm{d}$)} 
    \Else
      	\State \Do{\textsc{MergeSort}($\bm{y}', \bm{x}', \bm{q}', \bm{p}', \bm{d}$)}
      	\Comment{$\bm{z}' = (z_1,\dots,z_{\lfloor m/2 \rfloor})$}
      	 \State \Do{\textsc{MergeSort}($\bm{y}^*, \bm{x}^*, \bm{q}^*, \bm{p}^*, \bm{d}$)}
      	 \Comment{$\bm{z}^* = (z_{\lfloor m/2 \rfloor + 1},\dots,z_{m})$}
	     \State \Do{\textsc{Merge}($\bm{y}, \bm{x}, \bm{q}, \bm{p}, \bm{d}$)}
    \EndIf
    \Let{$\bm{y}$}{$\bm{x}$}
    \EndProcedure
  \end{algorithmic}
\end{algorithm}

\begin{algorithm}[H]
  \caption{\emph{Insertion sort} algorithm keeping the swap counts. It is inspired by the \texttt{C++} code underlying the \texttt{cor.fk} function of the \texttt{R} package \texttt{pcaPP} \citep{Filzmoser/Fritz/Kalcher:2018}.} \label{alg:insertion-sort}
  \begin{algorithmic}[1]
    \Require{$\bm{y}$: numeric vector of length $m$.} \Comment{observations}
    \Require{$\bm{q}, \bm{d}$: integer vectors of lengths $m,n$.} \Comment{ids and swap counts}
    \item[] ($m \leqslant n$, $\bm{q}$ is a subvector of $(1,\dots,n)$)
    \Statex
    \Procedure{InsertionSort}{$\bm{y}, \bm{q}$, $\bm{d}$}
    
		\For{$i = m-1,\dots,1$}
			\Let{$z$}{$y_i$}
	      	\Let{$\ell$}{$q_i$}
	      	\Let{$j$}{$i$}	      	
	      	\While{$j<m$}
	      		\If{$y_{j+1} < z$}
	          		\Let{$y_j$}{$y_{j+1}$}
	          		\Let{$q_j$}{$q_{j+1}$} \Comment{C1: new operation tracking (original) index}
	          		\Let{$d_{q_{j}}$}{$d_{q_{j}} + 1$} \Comment{C2: now updates individual swap counts}
	          		\Let{$j$}{$j+1$}
				\EndIf          		
	      	\EndWhile
          	\Let{$y_{j}$}{$z$}
          	\Let{$q_{j}$}{$\ell$} \Comment{see C1}
          	\Let{$d_{q_{j}}$}{$d_{q_{j}} + (j - i)$} \Comment{see C2}
		\EndFor        
    \EndProcedure
  \end{algorithmic}
\end{algorithm}

\begin{algorithm}[H]
\singlespace
  \caption{\emph{Merge} algorithm keeping the swap counts. It is inspired by the \texttt{C++} code underlying the \texttt{cor.fk} function of the \texttt{R} package \texttt{pcaPP} \citep{Filzmoser/Fritz/Kalcher:2018}.} \label{alg:merge-sort-single}
  \begin{algorithmic}[1]
    \Require{$\bm{y}, \bm{x}$: numeric vectors of lengths $m, m$.}  \Comment{observations and storage}
    \Require{$\bm{q}, \bm{p}, \bm{d}$: integer vectors of lengths $m, m, n$.} \Comment{ids, storage and swap counts}
    \item[] ($1 < m \leqslant n$, $\bm{q}$ is a subvector of $(1,\dots,n)$)

    \Statex
    \Procedure{Merge}{$\bm{y}, \bm{x}, \bm{q}, \bm{p}, \bm{d}$}
      \Let{$m'$}{$\lfloor m/2 \rfloor$}
      \Let{$m^*$}{$m - m'$}
      \Let{$i,j$}{$1$}
      \While{$i \leqslant m' \textrm{ and } j \leqslant m^*$}
        \If{$y_{i} > y_{m'+j}$}
        	\Let{$x_{i+j-1}$}{$y_{m'+j}$}
        	\Let{$p_{i+j-1}$}{$q_{m'+j}$} \Comment{C1: new operation tracking (original) index}
        	\Let{$d_{q_{m'+j}}$}{$d_{q_{m'+j}} + (m' - i + 1)$} \Comment{C2: now updates individual swap counts}
        	\Let{$j$}{$j + 1$}
        \Else
        	\Let{$x_{i+j-1}$}{$y_{i}$}
        	\Let{$p_{i+j-1}$}{$q_{i}$} \Comment{see C1}
        	\Let{$d_{q_{i}}$}{$d_{q_{i}} + (j - 1)$} \Comment{see C2}
        	\Let{$i$}{$i + 1$}
        \EndIf
      \EndWhile
      
      \If{$i < m'$}
      		\Let{$(x_{i}, \dots, x_{m'})$}{$(y_{i}, \dots, y_{m'})$}
      		\Let{$(p_{i}, \dots, p_{m'})$}{$(q_{i}, \dots, q_{m'})$} \Comment{see C1}
      		\Let{$(d_{q_{i}}, \dots, d_{q_{m'}})$}{$(d_{q_{i}}, \dots, d_{q_{m'}}) + m^*$} \Comment{see C2}
      \Else
      		\Let{$(x_{m'+j},\dots,x_{m})$}{$(y_{m'+j},\dots,y_{m})$}
      		\Let{$(p_{m'+j},\dots,p_{m})$}{$(q_{m'+j},\dots,q_{m})$} \Comment{see C1}
      \EndIf
    \EndProcedure
  \end{algorithmic}
\end{algorithm}

\begin{algorithm}[H]
  \caption{\emph{Divide-and-conquer} algorithm for computing $(\bm{d}^{(k)})_{k=2}^p$ from Section~3.} \label{alg:dac-within}
  \begin{algorithmic}[1]
    \Require{$\bm{X}$: integer matrix of dimensions $n \times p$.} \Comment{observations (as columnwise ranks)}
    \Require{$\bm{D}$: integer matrix of dimensions $n \times (p-1)$.}
    \Comment{discordance counts}
    \Require{$\bm{i}$: integer vector of length $n_0$.}
	\Comment{index vector}
    \Require{$k$: integer.} 
	\Comment{dimension considered}
    \Require{$n_*$: integer.} 
	\Comment{user-specified threshold}
    \item[] ($n_0 \leqslant n$, $\bm{i}$ is a subvector of $(1,\dots,n)$)
     \vspace{.25cm}    
     \item[]  \textsc{DivideAndConquer*} is given by Algorithm~\ref{alg:dac-between}. The remaining (simple) functions are available as part of the \texttt{C++} scripts also provided as online supplementary material. They are briefly described below.
     
     \vspace{.15cm}    
     \item[] $\bullet$ \textsc{Conquer}($\bm{X}, \bm{D}, \bm{i}, k$) naively computes the discordances within $(\bm{X}_i)_{i \in \bm{i}}$, and then sorts $\bm{i}$ so that $(X_{i,k+1})_{i \in \bm{i}}$ is sorted.
    \vspace{.15cm}    
    \item[] $\bullet$ \textsc{Merge*}($\bm{X}, \bm{D}, \bm{i}_1, \bm{i}_2, k$) is a slight variation of \textsc{Merge} (Algorithm~\ref{alg:merge-sort-single}).
    \vspace{.15cm}    
    \item[] $\bullet$ \textsc{FindSplitSort}($\bm{X}, \bm{D}, \bm{i}_1, \bm{i}_2, k$) is also a slight variation of \textsc{Merge} (Algorithm~\ref{alg:merge-sort-single}). It leaves $\bm{i}_1$ and $\bm{i}_2$ intact, and outputs $\bm{j} \in \mathbb{N}^2$, which records the optimal splits for $\bm{i}_1$ and $\bm{i}_2$ (see Section~3), and the permutation $\bm{i}'$ of $(\bm{i}_1,\bm{i}_2)$ for which $(X_{ik})_{i\in \bm{i}'}$ is sorted.

    \Statex
    \Procedure{DivideAndConquer}{$\bm{X}, \bm{D}, \bm{i}, k, n_*$}
    \\
    		\If{$n_0 < n_*$}
    			\Do{\textsc{Conquer}($\bm{X}, \bm{D}, \bm{i}, k$) and terminate.}
    		\Else
    			\State \Do{\textsc{DivideAndConquer}($\bm{X}, \bm{D}, \bm{i}_1, k, n_*$)}		
			\Comment{$\bm{i}_1 = (i_1,\dots,i_{m})$, $m=\lfloor n_0/2 \rfloor$}
    			\State \Do{\textsc{DivideAndConquer}($\bm{X}, \bm{D}, \bm{i}_2, k, n_*$)}
			\Comment{$\bm{i}_2 = (i_{m+1},\dots,i_{n_0})$}
		\\	
		\If{$k = p$}
			\Do{\textsc{Merge*}($\bm{X}, \bm{D}, \bm{i}_1, \bm{i}_2, k$) and terminate.}
		\EndIf
		\\
		\Let{$(\bm{j}, \bm{i}')$}{\textsc{FindSplitSort}($\bm{X}, \bm{D}, \bm{i}_1, \bm{i}_2, k$)}
		\For{$i \in \bm{i}_1^2$}
		\Comment{$\bm{i}_1^2 = (i_{j_1+1},\dots,i_{m})$}
		\Let{$D_{i,k-1}$}{$D_{i,k-1} + j_2$}
		\EndFor
		\For{$i \in \bm{i}_2^1$}
		\Comment{$\bm{i}_2^1 = (i_{m+1},\dots,i_{m+1+j_2})$}
		\Let{$D_{i,k-1}$}{$D_{i,k-1} + m - j_1$}
		\EndFor
		\\
    		\State \Do{\textsc{DivideAndConquer*}($\bm{X}, \bm{D}, \bm{i}_1^1, \bm{i}_2^1, k, 0, n_*$)}		
		\Comment{$\bm{i}_1^1 = (i_{1},\dots,i_{j_1})$}    		
		\State \Do{\textsc{DivideAndConquer*}($\bm{X}, \bm{D}, \bm{i}_1^2, \bm{i}_2^2, k, 0, n_*$)}
		\Comment{$\bm{i}_2^2 = (i_{j_2+1},\dots,i_{n_0})$}
		\\
		\If{$k < p$}
			\Do{\textsc{DivideAndConquer*}($\bm{X}, \bm{D}, \bm{i}_1^1, \bm{i}_2^2, k, 1, n_*$)}		
		\EndIf
		\\
		\Let{$\bm{i}$}{$\bm{i}'$}
    		
    	\EndIf    		
  \EndProcedure
  \end{algorithmic}
\end{algorithm}

\begin{algorithm}[H]
  \caption{\emph{Divide-and-conquer} algorithm for computing between-set discordances.} \label{alg:dac-between}
  \begin{algorithmic}[1]
    \Require{$\bm{X}$: integer matrix of dimensions $n \times p$.} \Comment{observations}
    \Require{$\bm{D} $: integer matrix of dimensions $n \times (p-1)$.}
    \Comment{discordance counts}
    \Require{$\bm{i}_1, \bm{i}_2$: integer vectors of lengths $n_1, n_2$.}
	\Comment{index vectors}
    \Require{$k$: integer.} 
	\Comment{dimension considered}
    \Require{$b$: either $0$ or $1$.} 
	\Comment{triggers extra sorting step}
    \Require{$n_*$: integer.} 
	\Comment{user-specified threshold}
    \item[] ($n_1,n_2 \leqslant n$, $\bm{i}_1$ and $\bm{i}_2$ are subvectors of $(1,\dots,n)$, $\bm{X}$ consists of columnwise ranks)
     \vspace{.25cm}    
     \item[]  The simple functions described below are available as part of the \texttt{C++} scripts also provided as online supplementary material.
     
     \vspace{.15cm}    
     \item[] $\bullet$ \textsc{Conquer*}($\bm{X}, \bm{D}, \bm{i}_1, \bm{i}_2, k$) naively computes the discordances between $(\bm{X}_i)_{i \in \bm{i}_1}$ and $(\bm{X}_i)_{i \in \bm{i}_2}$.
    \vspace{.15cm}    
    \item[] $\bullet$ \textsc{Sort}($\bm{i}, \bm{X}, k$) permutes $\bm{i}$ so that $(X_{ik})_{i\in \bm{i}}$ is sorted.
    \vspace{.15cm}    
    \item[] $\bullet$ \textsc{Merge**}($\bm{X}, \bm{D}, \bm{i}_1, \bm{i}_2, k$) is a slight variation of \textsc{Merge} (Algorithm~\ref{alg:merge-sort-single}).
    \vspace{.15cm}    
    \item[] $\bullet$ \textsc{FindSplit}($\bm{X}, \bm{D}, \bm{i}_1, \bm{i}_2, k$) is also a slight variation of \textsc{Merge} (Algorithm~\ref{alg:merge-sort-single}). It leaves $\bm{i}_1$ and $\bm{i}_2$ intact, and outputs $\bm{j} \in \mathbb{N}^2$, which records the optimal splits for $\bm{i}_1$ and $\bm{i}_2$.

    \Statex
    \Procedure{DivideAndConquer*}{$\bm{X}, \bm{D}, \bm{i}_1, \bm{i}_2, k, b, n_*$}
    \\
    		\If{$\sqrt{n_1n_2} < n_*$ or $\min(n_1,n_2) = 1$}
    		
    			\Do{\textsc{Conquer*}($\bm{X}, \bm{D}, \bm{i}_1, \bm{i}_2, k$) and terminate.}
    		\Else
    		
		\\	
		\If{$b = 1$}
			\State \Do{\textsc{Sort}($\bm{i}_1, \bm{X}, k$)}
			\State \Do{\textsc{Sort}($\bm{i}_2, \bm{X}, k$)}
		\EndIf

		\\	
		\If{$k = p$}
			\Do{\textsc{Merge**}($\bm{X}, \bm{D}, \bm{i}_1, \bm{i}_2, k$) and terminate.}
		\EndIf
		\\
		\Let{$\bm{j}$}{\textsc{FindSplit}($\bm{X}, \bm{D}, \bm{i}_1, \bm{i}_2, k$)}
		\For{$i \in \bm{i}_1^2$}
		\Comment{$\bm{i}_1^2 = (i_{1,j_1+1},\dots,i_{1,n_1})$}
		\Let{$D_{i,k-1}$}{$D_{i,k-1} + j_2$}
		\EndFor
		\For{$i \in \bm{i}_2^1$}
		\Comment{$\bm{i}_2^1 = (i_{2,1},\dots,i_{2,j_2})$}				\Let{$D_{i,k-1}$}{$D_{i,k-1} + n_1 - j_1$}
		\EndFor
		\\
    		\State \Do{\textsc{DivideAndConquer*}($\bm{X}, \bm{D}, \bm{i}_1^1, \bm{i}_2^1, k, 0, n_*$)}		
		\Comment{$\bm{i}_1^1 = (i_{1,1},\dots,i_{1,j_1})$}
		\State \Do{\textsc{DivideAndConquer*}($\bm{X}, \bm{D}, \bm{i}_1^2, \bm{i}_2^2, k, 0, n_*$)}
		\Comment{$\bm{i}_2^2 = (i_{2,j_2+1},\dots,i_{2,n_2})$}
		\\
		\If{$k < p$}
			\Do{\textsc{DivideAndConquer*}($\bm{X}, \bm{D}, \bm{i}_1^1, \bm{i}_2^2, k, 1, n_*$)}		
		\EndIf
    		
    	\EndIf    		
  \EndProcedure
  \end{algorithmic}
\end{algorithm}

\newpage
\section{Runtime analysis}
\renewcommand{\theequation}{\thesection.\arabic{equation}}
\normalsize
\spacingset{1.45}

For convenience, we focus on the case $p \geqslant 3$.
Given sets $\mathcal{I}$ and $\mathcal{J}$, let $T(\mathcal{I})$ be the runtime of Algorithm~A.4 for computing $\bm{D}_{\mathcal{I}}$, and $T_k(\mathcal{I},\mathcal{J})$ and $T_k^*(\mathcal{I},\mathcal{J})$ be that of Algorithm~A.5 for computing $(D_{\mathcal{I}\mathcal{J}}^{(r)})_{r=k}^p$, where $k \in \{2,\dots,p\}$, where the asterisk indicates an extra sorting step.
With $\Theta(n)$ representing the time required for spliting and merging index sets, we have, for $k < p$, $\mathcal{I} = \{1,\dots,n\}$ with $n$ large, and other sets as in Section~3,
\vspace{-.4cm}
\begin{align}
	T(\mathcal{I}) &= \sum_{\delta=0,1} T(\mathcal{I}_{\delta}) + \sum_{\gamma=0,1} T_2(\mathcal{I}_{0}^\gamma,\mathcal{I}_{1}^\gamma) + T_{3}^*(\mathcal{I}_{0}^0,\mathcal{I}_{1}^1) + \Theta(n)\;,\label{eq:time-within} \\
	T_k(\mathcal{I}_0,\mathcal{I}_1) &= \sum_{\gamma=0,1} T_k(\mathcal{I}_{0}^\gamma,\mathcal{I}_{1}^\gamma) + T_{k+1}^*(\mathcal{I}_{0}^0,\mathcal{I}_{1}^1) + \Theta(n)\;, \label{eq:time-between}
\end{align}
\vspace{-.8cm}

\noindent and $T_k^*(\mathcal{I}_{0}^\gamma,\mathcal{I}_{1}^\gamma) = T_k(\mathcal{I}_{0}^\gamma,\mathcal{I}_{1}^\gamma) + O(n \log_2 n)$, accounting for the extra sorting step.
The case $k=p$ is similar, with the exception that $T_{k+1}^*(\mathcal{I}_{0}^0,\mathcal{I}_1^1)$ drops out of \eqref{eq:time-between}.
We now derive explicit expressions for $T(\mathcal{I})$ in two specific scenarios where the various sets are easily tractable, and we then build on these to conclude that $T(\mathcal{I}) = O(n \log^p n)$ in general.

\emph{Co-monotonicity}. As a first scenario, suppose that $n=2^K$ for some $K \in \mathbb{N}$ and that all pairs of observations are concordant.
In this case, $\mathcal{I} = \mathcal{I}_{0}^0 \cup \mathcal{I}_{1}^1$ and $\mathcal{I}_{0}^1=\mathcal{I}_{1}^0=\emptyset$, and so $T_k(\mathcal{I}_{0}^\gamma,\mathcal{I}_{1}^\gamma) = \Theta(1)$ for both $\gamma \in \{0,1\}$ and all $k$.
To deduce $T_{3}^*(\mathcal{I}_{0}^0,\mathcal{I}_{1}^1)$ in \eqref{eq:time-within}, note that, provided that the extra sorting step is handled in $\Theta(n)$ operations for sorted vectors, $T_{k}^*(\mathcal{I}_{0}^0,\mathcal{I}_{1}^1) = T_{k+1}^*(\mathcal{I}_{0}^0,\mathcal{I}_{1}^1) + \Theta(n)$ for $k \in \{3,\dots,p-1\}$, where $T_{p}^*(\mathcal{I}_{0}^0,\mathcal{I}_{1}^1) = \Theta(n)$.
Consequently, $T_{k}^*(\mathcal{I}_{0}^0,\mathcal{I}_{1}^1) = \Theta(n)$ for all $k \in \{3,\dots,p\}$.
Plugging this in \eqref{eq:time-within} and noting that $T(\mathcal{I}_0) = T(\mathcal{I}_1)$ gives $T(\mathcal{I}) = 2 T(\mathcal{I}_0) + \Theta(n)$, where $|\mathcal{I}_0| = n/2$.
A direct application of the Master Theorem \cite[Theorem~4.1]{Cormen/al:2014} shows that $T(\mathcal{I}) = \Theta(n \log_2 n)$. 

\emph{Perfect uniformity}. Suppose now that $n = 2^{Kp}$ for some $K \in \mathbb{N}$ and that the data is evenly distributed on $\{1,\dots,n\}^p$ (ignoring the existence of ties).
In this case, $|\mathcal{I}_\delta|=n/2$ and $|\mathcal{I}_\delta^\gamma|=n/4$ for any $\delta,\gamma \in \{0,1\}$.
Consider first $T_k^*(\mathcal{I}_0,\mathcal{I}_1)$ and recall that $T_{p}^*(\mathcal{I}_0,\mathcal{I}_1)$ is at worst $\Theta(n \log_2 n)$, as it involves sorting each set and then merging them.
Again from the Master Theorem \citep[see][]{Bentley/Haken/Saxe:1980}, we get $T_{p-1}^*(\mathcal{I}_0,\mathcal{I}_1)=2 T_{p-1}(\mathcal{I}_0^\gamma,\mathcal{I}_1^\gamma)+\Theta(n \log_2 n)=\Theta(n \log_2^2 n)$ and, proceeding recursively, $T_{k+1}^*(\mathcal{I}_0,\mathcal{I}_1)= \Theta(n \log_2^{p-k} n)$. 
Reasoning similarly, we obtain $T_{k}(\mathcal{I}_0,\mathcal{I}_1) = \Theta(n \log_2^{p-k+1} n)$.
Since $T_2(\mathcal{I}_0^\gamma,\mathcal{I}_1^\gamma) = \Theta(n \log_2^{p-1} n)$ dominates the rightmost sum in \eqref{eq:time-within}, the equation reduces to $T(\mathcal{I}) = 2T(\mathcal{I}_0) + \Theta(n \log_2^{p-1} n)$, and a final application of the Master Theorem shows that $T(\mathcal{I}) = \Theta(n \log_2^p n)$.

\emph{General case}. Let $n_{\delta\gamma} = |\mathcal{I}_\delta^\gamma|$, $\delta,\gamma \in \{0,1\}$.
To bound $T(\mathcal{I})$ for arbitrary data, let us first find a (potentially impossible) configuration of the observations that maximises the amount of work required in \eqref{eq:time-within}.
To this end, it suffices to independently bound $N_1 = n_{00}n_{11}$ and $N_2 = n_{00}n_{10}+n_{01}n_{11}$; $N_1$ is the number of pairs for which only a partial conclusion (partial concordance) is reached, and $N_2$ is the number of pairs for which no conclusion is reached.
Clearly, $N_1$ is maximised when $n_{00}=n_{11}=n/2$.

To bound $N_2$, first recall that the sets $\mathcal{I}_{\delta}^\gamma$ are constructed so as to maximise $\beta := n_{00}n_{11} + n_{01}n_{10}$, where $n_{\delta 0}+n_{\delta 1}=n/2$, $\delta \in \{0,1\}$.
This means in particular that, for any $\gamma \in \{0,1\}$, getting both $n_{0\gamma}, n_{1\gamma} > n/4$ is impossible.
To see this, suppose without loss of generality that $n_{00}, n_{10} > n/4$, which implies $n_{00} > n_{01}$ and $n_{10} > n_{11}$, and consider moving the split value ($s$ in Section~3) down so that one and only one observation, say the $i$th, is affected.
If $i \in \mathcal{I}_0$ (in which case it moved from $\mathcal{I}_0^0$ to $\mathcal{I}_0^1$), then the new split criterion is $(n_{00}-1)n_{11} + (n_{01}+1)n_{10} = n_{00}n_{11} + n_{01}n_{10} - n_{11} + n_{10} > \beta$, since $n_{10} > n_{11}$ by assumption, and $\beta$ is sub-optimal.
Similarly, if $i \in \mathcal{I}_1$, then the new split criterion is $n_{00}(n_{11}+1) + n_{01}(n_{10}-1) = n_{00}n_{11} + n_{01}n_{10} + n_{00} - n_{01} > \beta$.
Thus, under the constraint that $\beta$ is optimal, $N_2$ is bounded by $2(n/4)(n/4)$, which occurs when $n_{\delta\gamma}=n/4$, $\delta,\gamma \in \{0,1\}$.

Given the above considerations, we can bound $T(\mathcal{I})$ by considering, instead of \eqref{eq:time-within} and \eqref{eq:time-between}, the new runtime equations
\vspace{-.4cm}
\begin{align*}
	T(\mathcal{I}) &= 2 T(\mathcal{I}_{0}) + 2 T_2(\mathcal{I}_{0}^0,\mathcal{I}_{1}^0) + T_{3}^*(\mathcal{J}_{0}^0,\mathcal{J}_{1}^1) + \Theta(n)\;,\\
	T_k(\mathcal{I}_0,\mathcal{I}_1) &= 2 T_k(\mathcal{I}_{0}^0,\mathcal{I}_{1}^0) + T_{k+1}^*(\mathcal{J}_{0}^0,\mathcal{J}_{1}^1) + \Theta(n)\;,
\end{align*}
\vspace{-1.25cm}

\noindent where $|\mathcal{I}_0^0| = |\mathcal{I}_1^0| = n/4$ and $|\mathcal{J}_0^0|=|\mathcal{J}_0^0|=n/2$.
Now, recall that, at worst, $T_p^*(\mathcal{J}_0^0,\mathcal{J}_1^1) = \Theta(n \log_2 n)$, and that, from the perfectly uniform case, $T_{p}(\mathcal{I}_0^0,\mathcal{I}_1^0) = \Theta(n \log_2 n)$.
Further note that $T_p^*(\mathcal{J}_0^0,\mathcal{J}_1^1)$ is of the same order as $T_p^*(\mathcal{I}_0^0,\mathcal{I}_1^1)$ under perfect uniformity; doubling the size of the sets does not significantly affect the runtime.
Proceeding recursively, we also recover $T_{k+1}^*(\mathcal{I}_0,\mathcal{I}_1)= \Theta(n \log_2^{p-k} n)$ and $T_{k}(\mathcal{I}_0,\mathcal{I}_1) = \Theta(n \log_2^{p-k+1} n)$ as in the perfectly uniform case, from which we again deduce that $T(\mathcal{I}) = \Theta(n \log_2^p n)$.
Now, the use of $\Theta( \cdot )$ here is valid only for data which cannot in reality occur ($\mathcal{I}_0^0$ must be the same as $\mathcal{J}_0^0$, but $|\mathcal{I}_0^0| \neq |\mathcal{J}_0^0|$), and so $T(\mathcal{I}) = O(n \log_2^p n)$. As we have shown, this bound is attained in the perfectly uniform case.

\newpage
\section{Supplemental material for Section~4}  \label{app:benchmark}
\normalsize
\spacingset{1}

The experiments were executed on a Dell Precision 5820 machine (memory: 194GiB, processor: Intel(R) Xeon(R) W-2145 CPU @ 3.70GHz $\times$ 16, operating system: 64-bit Ubuntu 20.04.4 LTS).

\begin{figure}[H]
\singlespace
\centering
\includegraphics[width=.8\textwidth]{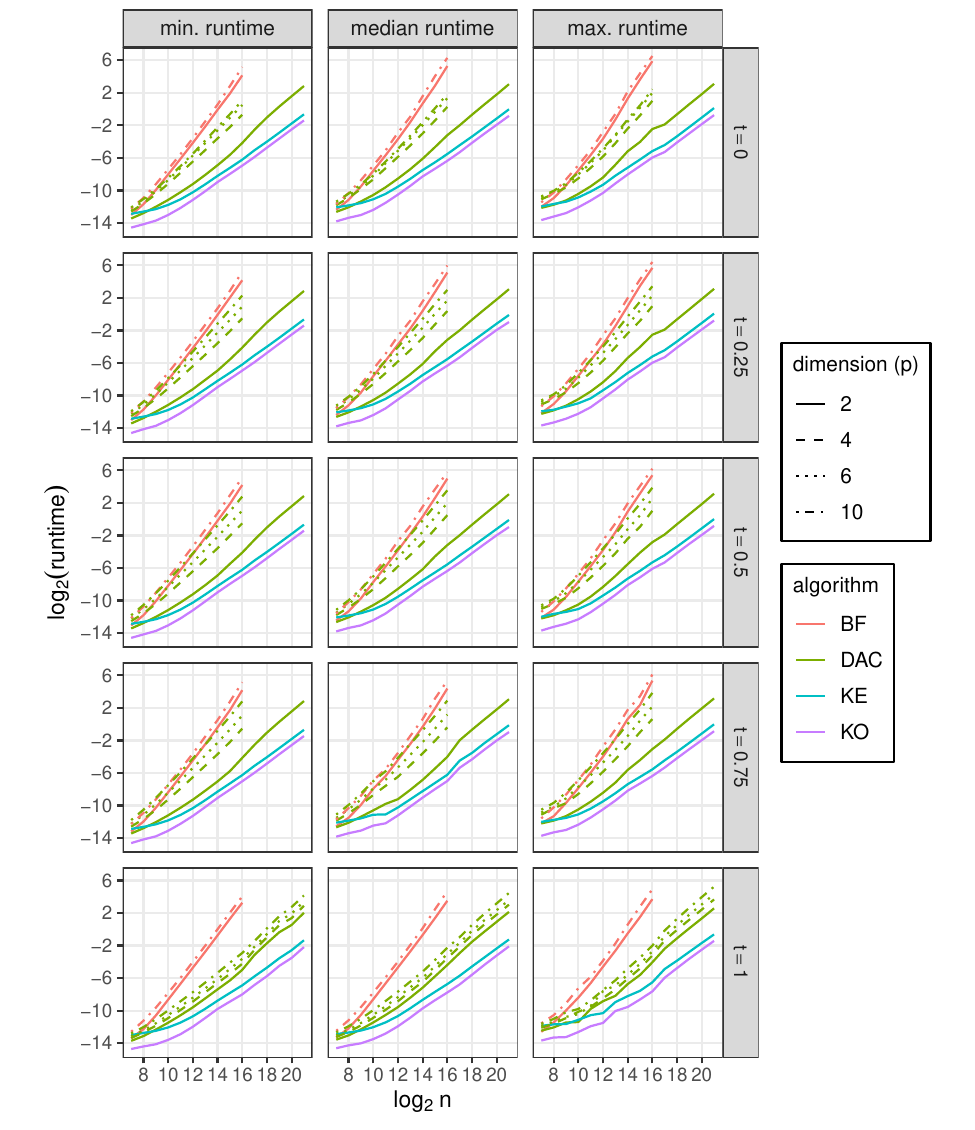}
\caption{\small Minimum, median and maximum runtimes, over $100$ datasets, of Knight's original algorithm (KO, Section~2.2), the extension (KE) from Section~2.3, the divide-and-conquer algorithm (DAC) from Section~3 and their naive alternative (BF) for computing $\bm{\hat\tau}_p$ and (except for KO) $\bm{\hat\Sigma}_p$ for each $p \in \{2,4,6,10\}$ (when applicable).
The data are equicorrelated normal random vectors with corresponding correlation $\rho_t = \sin(\pi t/2)$, $t \in \{i/4\}_{i=0}^4$.
Given a dataset, the computations with KO and KE were replicated $100$ times; those with DAC $50$ times whenever $t=1$ or $p=2$, and $10$ times otherwise; and those with BF $50$ times when $n \leqslant 2^{12}$ and $3$ times otherwise.} \label{fig:bench-full}
\end{figure}

\begin{figure}[H]
\singlespace
\centering
\includegraphics[width=1\textwidth]{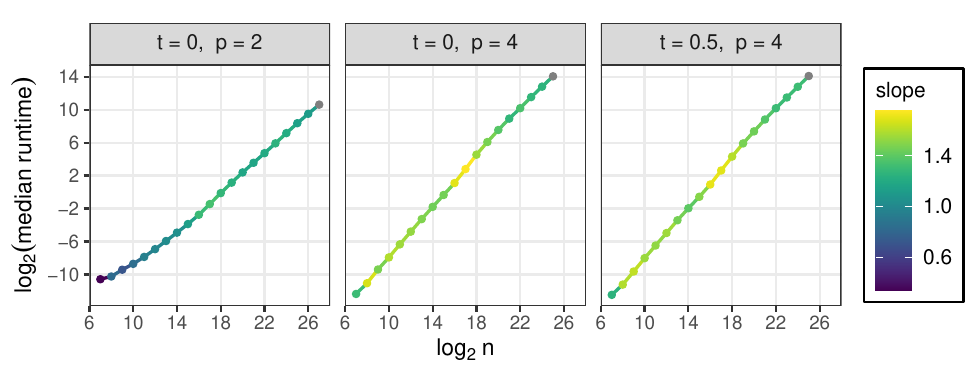}
\caption{\small Median runtimes of the divide-and-conquer algorithm (DAC) from Section~3 computing $\bm{\hat\tau}_p$ and $\bm{\hat\Sigma}_p$, $p \in \{2,4\}$.
The data are equicorrelated normal random vectors with corresponding correlation $\rho_t = \sin(\pi t/2)$.
Datasets were generated independently for each values of $n \in \{2^k : k=7,\dots,K\}$; $K=27$ for $p=2$ and $K=25$ for $p=4$.
For $p=2$, five datasets for each value of $n$ were generated, and only a single one was generated for $p=4$; each computation was replicated $5$ times.
The rightmost segment in each plot has a slope of ${\sim}1.11$, ${\sim}1.25$, and ${\sim}1.3$, respectively (from left to right).
} \label{fig:extra}
\end{figure}

\newpage
\section{Multi-lag serial dependence} \label{app:multi-lag}
\normalsize
\spacingset{1.45}

\subsection{Definition and inference} \label{sec:ml-def}

In this Appendix, we investigate how $\hat\tau_p$ can unveil interesting temporal characteristics of observed phenomena.
Moving in this direction, \cite{Ferguson/Genest/Hallin:2000} defined a $\hat\tau_2$-based autocorrelation measure, which they used to test against serial dependence in univariate time series.
In this section, we go beyond the pairwise case and define a $\hat\tau_p$-based autocorrelation measure that captures the temporal persistence of concordance.
Let $(X_i)_{i=1}^{n+p-1}$ denote the sequence under study and assume, for simplicity, that the underlying process is a stationary, finite-order autoregressive process.
More comprehensive treatments can be found in \cite{Sen:1972}, \cite{Yoshihara:1976}, and \cite{Harel/Puri:1989}.
To quantify persistence, we construct our dataset $(\bm{X}_i)_{i=1}^n$ by letting $\bm{X}_i = (X_{i+k})_{k=0}^{p-1}$, enabling us to compute $\bm{\hat\tau}_p := (\hat\tau_k)_{k=2}^{p}$ using the divide-and-conquer algorithm outlined in Section~3.
Here, the index $k$ of $\hat\tau_k$ stands for the width of the time window considered, and high values of $\hat\tau_k$ indicate that the process under study is in some sense stable over such windows.

The limiting behaviour of $\bm{\hat\tau}_p$ also follows from the results of \cite{Sen:1972} for U-statistics.
To be specific, let $\bm{\tau}_p$ be the expected value of $\bm{\hat\tau}_p$ given two independent copies of $\bm{X}_1$ and, for all integers $j \geqslant 0$, denote by $\mathbf{Z}_j$ the corresponding matrix extension of $\zeta_{j}$ (as defined between (1) and (2)).
Then, from Theorem~1 of \cite{Sen:1972}, $\sqrt{n}(\bm{\hat\tau}_p - \bm{\tau}_p) \rightsquigarrow \mathcal{N}(\bm{0}, \bm{\Sigma}_p)$, with $\bm{\Sigma}_p := 4 \mathbf{Z}_0 + 4 \sum_{j=1}^\infty(\mathbf{Z}_j + \mathbf{Z}_j^\top)$ as $n \to \infty$.
We now generalise this methodology to allow for the joint analysis of two dependent time series with varying sample sizes.

Consider two time series and their corresponding $n_i \times p$ datasets ($i=1,2$) of complete observations, constructed as explained above.
Further let $n_{12}$ and $n$ denote the number of shared and unique timestamps, respectively ($n = n_1 + n_2 - n_{12}$).
To assess whether the discrepancy between the corresponding measures, say $\bm{\hat\tau}_p^1$ and $\bm{\hat\tau}_p^2$, is significant, let us examine the statistic $\mathcal{D} = \sqrt{n} (\bm{\hat\tau}_p^1 - \bm{\hat\tau}_p^2)$ under the asymptotic regime where $n/n_i \to \gamma_i$ and $(nn_{12})/(n_1n_2) \to \gamma_{12}$ as $n \to \infty$ for some $\gamma_i,\gamma_{12} > 0$.
Then, as before, there exist matrices $\bm{\Sigma}_p^i$ ($i=1,2$) such that $n \Var(\bm{\hat\tau}_p^i) = (n/n_i) n_i \Var(\bm{\hat\tau}_p^i) \to_{\mathbb{P}} \gamma_i \bm{\Sigma}_p^i$, and a matrix $\bm{\Sigma}_p^{12}$ similarly defined for $\Cov(\bm{\hat\tau}_p^1,\bm{\hat\tau}_p^2)$.
In particular, $\Var(\mathcal{D}) \to_{\mathbb{P}} \bm{\Sigma}_* := \gamma_1 \bm{\Sigma}_p^1 + \gamma_2 \bm{\Sigma}_p^2 - \gamma_{12}\{\bm{\Sigma}_p^{12} + (\bm{\Sigma}_p^{12})^\top\}$.
A consistent estimator $\bm{\hat\Sigma}_*$ of $\bm{\Sigma}_*$ is obtained by replacing each of its constituents by their empirical analogues.
It allows us to test the hypothesis $\bm{\tau}_p^1 = \bm{\tau}_p^2$ using, \emph{e.g.}, the Mahalanobis-type statistic $\mathcal{D}^\top \bm{\hat\Sigma}_*^{-1} \mathcal{D}$, whose limiting distribution is $\chi_{p}^2$ as $n\to\infty$ (provided that $\bm{\Sigma}_*$ is positive definite).

\subsection{Illustrative application: temperature in Ontario}  \label{sec:application}

\begin{figure}[!b]
\centering
\includegraphics[scale=.88]{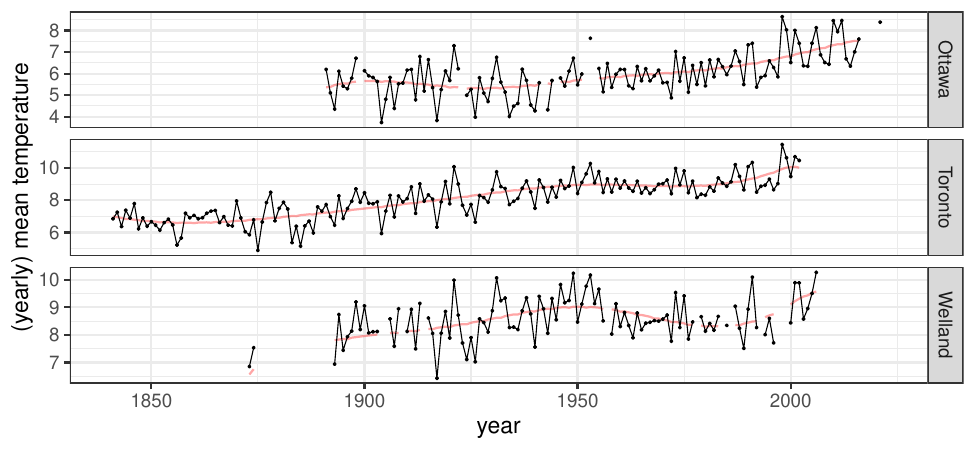}
\caption{Yearly mean temperatures (in \SI{}{\celsius}, complete years only) measured in Ottawa, Toronto and Welland. The curves in red are the fitted (conditional) means, as described in Section~5.2, also averaged yearly.} \label{fig:fit-yearly}
\end{figure}

\begin{figure}[!t]
\centering
\includegraphics[scale=.88]{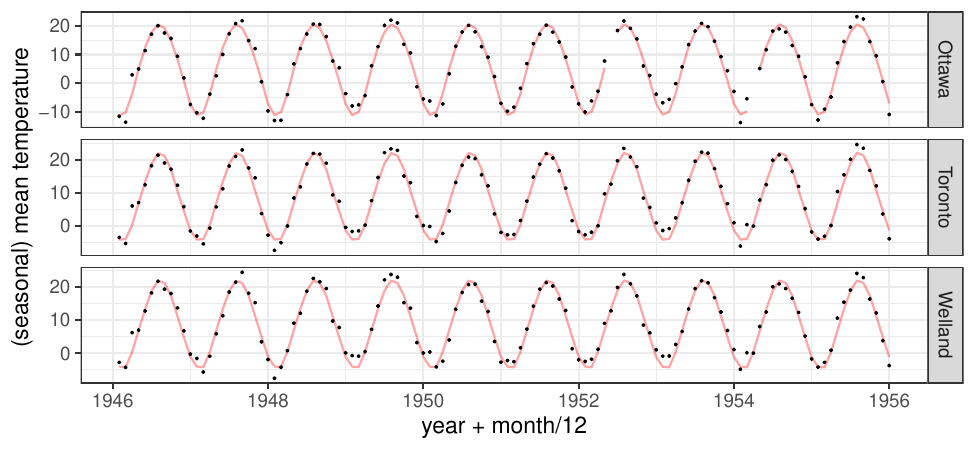}
\caption{Monthly mean temperatures (in \SI{}{\celsius}, for the years 1946 to 1955, inclusively) measured in Ottawa, Toronto and Welland. The curves in red are the fitted (conditional) means, as described in Section~5.2, also averaged montly.} \label{fig:fit-seas}
\end{figure}

\textbf{Data.}
As illustrative example, we analyse daily mean temperatures measured in the Ontarian cities of Ottawa (from 1889-11-01 to 2023-09-07, 48,732 obs.), Toronto (from 1840-03-01 to 2003-06-30, 59,651 obs.)~and Welland (from 1872-10-01 to 2014-08-09, 47,620 obs.).~The data were downloaded from Weather Canada's website\footnote{\url{https://climate.weather.gc.ca/historical_data/search_historic_data_e.html}} using the \texttt{weathercan} package \citep{weathercan:2018} available in \texttt{R}.
The raw data are illustrated in Figures~\ref{fig:fit-yearly} and \ref{fig:fit-seas}.
Unsuprisingly, they reveal the presence of seasonal and long-term trends.

\begin{figure}[!t]
\centering
\includegraphics[scale=.88]{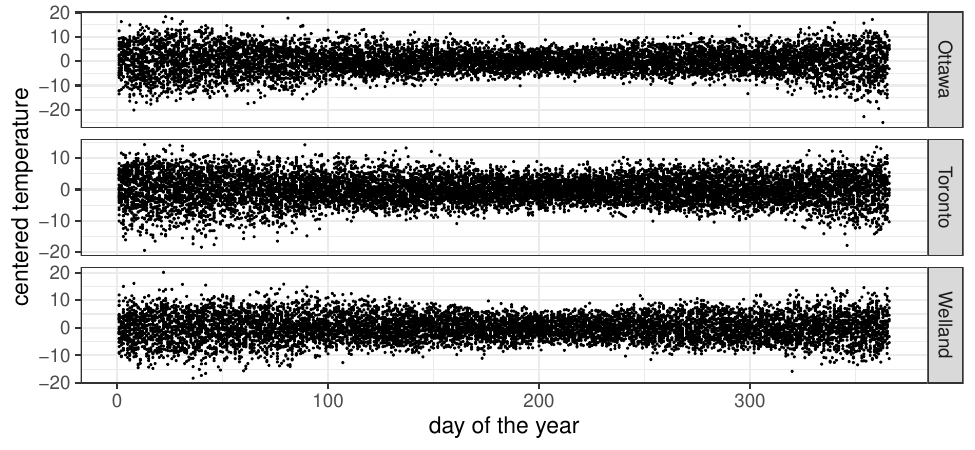}
\caption{Subset of the city-specific residuals (centered temperatures, in \SI{}{\celsius}) resulting from the least-squares fit described in Section~5.2, plotted against their corresponding day of the year.} \label{fig:residuals}
\end{figure}

\begin{figure}[!t]
\centering
\includegraphics[scale=.88]{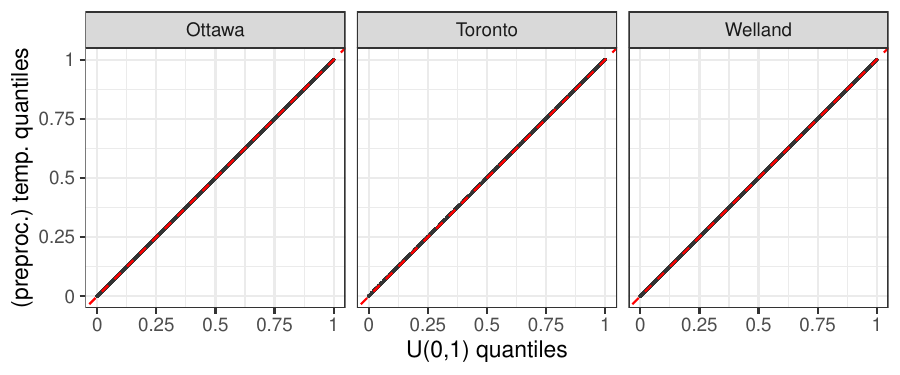}
\caption{City-specific qq-plots of the pseudo-observations upon which the analysis of Section~\ref{sec:application} is based against the uniform distribution. The preprocessing involved centering (see Figure~\ref{fig:fit-yearly}--\ref{fig:residuals}) and applying an empirical version of the probability integral transform.} \label{fig:pseudo-qq}
\end{figure}

\textbf{Preprocessing.}
To address the underlying trends, we centered the data by estimating their mean through least squares. The design matrices included sines and cosines with frequencies ranging from one to six cycles per year, as well as b-splines of degree three. The interior knots for the b-splines were placed on Jan.\ 1st of the years 1882 (except for Ottawa), 1925, 1953 and 1997. We further included interactions between the b-splines and the sines and cosines with a frequency of one cycle per year.
The resulting residuals, denoted $\epsilon_{ik}$, are shown in Figure~\ref{fig:residuals}, which suggests the presence of heteroskedasticity over the seasonal cycle.
To address this, we applied an empirical version of the probability integral transform: for a given city $k$ and each timestamp $i$ falling on day, say $t_{i}$, of the year, we computed $U_{ik} := (1/s) \sum_{j=1}^n \phi(t_j-t_i) \mathbbm{1}(\epsilon_{ik} \leqslant \epsilon_{jk})$, where $\phi$ is the Gaussian density (mean $0$, stand.~dev.~$2$) and $s = \sum_{j} \phi(t_j-t_i)$.
We based our analysis on these pseudo-observations, which we assume to follow a $U(0,1)$ distribution (see Figure~\ref{fig:pseudo-qq}).

\begin{figure}[t]
\centering
\includegraphics[scale=.88]{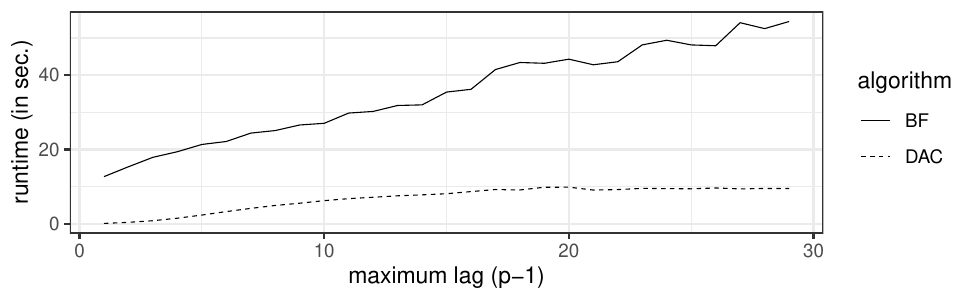}
\caption{Runtimes (in seconds), for each value of $p \in \{2,\dots,30\}$, of the divide-and-conquer (DAC) and brute force (BF) algorithms for computing $\bm{\hat\tau}_p$ and $\bm{\hat\Sigma}_p$, as described in Section~\ref{sec:ml-def}, underlying the Toronto temperature time series ($n=59,651$, see Section~\ref{sec:application}).} \label{fig:app-runtimes}
\end{figure}

\textbf{Runtimes.}
For comparison, we computed $\bm{\hat\tau}_p$ and $\bm{\hat\Sigma}_p$ for each value of $p \in \{2,\dots,30\}$ using both BF and DAC.
The resulting runtimes are provided in Figure~\ref{fig:app-runtimes}.
While the runtimes of both algorithms rapidly increase with $p$ for smaller values of the latter, that of DAC is much more stable for large $p$, as it benefits from the fact that $\hat\tau_\ell$ is negligible for large $\ell$.
DAC outperformed BF for all values of $p$ considered, with an efficiency ratio starting from $\sim$164 for $p=2$, decreasing to $\sim$4 at $p=15$, and then slowly increasing onward, reaching $\sim$5.5 at $p = 30$. In this latter case, BF takes nearly a minute to run.

\begin{figure}[t]
\centering
\includegraphics[scale=1]{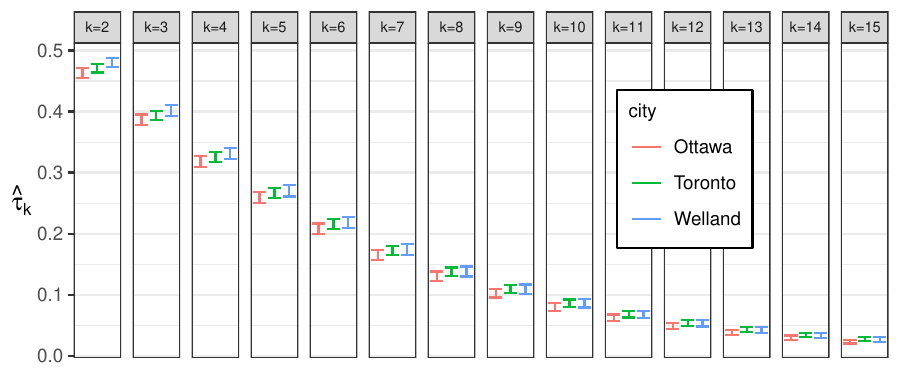}
\caption{Depiction of the vectors $(\hat\tau_k)_{k=2}^{15}$ quantifying the persistence of daily mean temperatures in Ottawa, Toronto and Welland, Ontario, computed as described in Section~\ref{sec:application}. Each set of error bars represents the $99\%$ pointwise confidence interval, symmetric around the point estimate, based on jackknife estimation.} \label{fig:tau-example}
\end{figure}

\textbf{Analysis.}
Figure~\ref{fig:tau-example} portrays the vector $\bm{\hat\tau}_{15}$ derived from each univariate series, along with their respective 99\% pointwise confidence intervals (with $m$ set to $20$ based on a visual analysis of the matrices $\mathbf{Z}_j$ of Section~\ref{sec:ml-def}).
The results align well with intuition. The high and slowly decaying (with respect to $k$) values suggest that temperature is relatively stable over time.
Moreover, the three curves are very similar, with Toronto's curve positioned between those of Welland and Ottawa, as one could expect based on their relative geographical positions.
The measures for Toronto and Welland, which are geographically closer to one another, exhibit a greater overall similarity, but due to the larger confidence intervals for Ottawa, we see more overlap between Ottawa and Toronto than Welland and Toronto for $k \in \{2,3\}$.
Owing to the large sample sizes considered and the resulting narrow confidence intervals, there appears to be substantial evidence to reject the hypothesis that the persitence of concordance at all three locations is identical, at least for time windows of width $k \in \{2,3\}$.

Using the statistic $\mathcal{D}$ defined in Section~\ref{sec:ml-def}, we compared the measures $\bm{\hat\tau}_p$ for Toronto with those for Welland and Ottawa. When focusing on $\bm{\hat\tau}_2 = \hat\tau_2$, the resulting p-values ($0.001$ for Toronto-Welland and $0.027$ for Toronto-Ottawa) indeed suggest a significant difference in serial dependence. Similar results for Toronto-Welland ensue for $\bm{\hat\tau}_3 = (\hat\tau_2, \hat\tau_3)$, but the test at level $0.05$ does not reject equality for Toronto-Ottawa (p-val: $0.09$) in this case. 
As more components are considered, the power of the test decreases, and for $\bm{\hat\tau}_{12}$ none of the two hypotheses is rejected at level $0.05$.

\textbf{Cautionary remarks.}
It must be noted that the above analysis is simplistic and meant to give insights into the behaviour of the computing algorithms with large datasets and to suggest potential uses for $\bm{\hat\tau}_p$.
In particular, it is implicitly assumed that the (preprocessed) series are stationary; a more reasonable approach might involve segmenting the analysis into season-specific ones.
The broad conclusions of the analysis, however, will likely stay unchanged.
Furthermore, the content of this section raises new questions about the behaviour of $\bm{\hat\tau}_p$ for $p$ large, since $\hat\tau_p$ and, more importantly, $\hat\sigma_p^2$ can tend to zero very quickly with $p$ (see Figure~\ref{fig:tau-example}).

\nocite{Mersmann:2021}

\singlespace
\bibliographystyle{apalike}

\end{document}